\documentclass[onecolumn,superscriptaddress,longbibliography,12pt]{revtex4-1}
\usepackage{graphicx}
\usepackage{amssymb}
\usepackage{amsmath}
\usepackage{epsfig}
\usepackage{color}
\usepackage{natbib}
\usepackage{mathtools}
\usepackage[colorlinks,linkcolor=blue,anchorcolor=blue,citecolor=blue,urlcolor=blue]{hyperref}
\usepackage[left]{lineno}
\usepackage{blindtext}
\usepackage{color, soul}
\usepackage{setspace}
\usepackage{fancyhdr}

\begin{document}

\title{Room-temperature Composite-pulses for Robust Diamond Magnetometry}
\author{Jing-Yan Xu}
\author{Yang Dong}
\author{Shao-Chun Zhang}
\author{Yu Zheng}
\author{Xiang-Dong Chen}
\affiliation{\textsf{CAS Key Lab of Quantum Information, University of Science and Technology of China, Hefei,
230026, P.R. China}}
\affiliation{\textsf{CAS Center For Excellence in Quantum Information and Quantum Physics, University of Science
and Technology of China, Hefei, 230026, P.R. China}}
\author{Wei Zhu}
\author{Guan-Zhong Wang}
\affiliation{\textsf{Hefei National Laboratory for Physical Science
at Microscale, and Department of Physics, University of Science and Technology of China, Hefei, Anhui 230026, P. R. China}}
\author{Guang-Can Guo}
\author{Fang-Wen Sun}
\email{fwsun@ustc.edu.cn}
\affiliation{\textsf{CAS Key Lab of Quantum Information, University of Science and Technology of China, Hefei,
230026, P.R. China}}
\affiliation{\textsf{CAS Center For Excellence in Quantum Information and Quantum Physics, University of Science
and Technology of China, Hefei, 230026, P.R. China}}
\date{\today}

\begin{abstract}
\textsf{The sensitivity of a practical quantum magnetometer is challenged by both inhomogeneous coupling between sensors and environment and errors in quantum control. Based on the physical criteria of modern quantum sensing, we present a robust and effective composite-pulse for high fidelity operation against inhomogeneous broadening of sensor and control errors via optimized modulation of control field. Such a technique is verified on nitrogen-vacancy (NV) center in diamond to keep almost perfect operation up to a detuning as larger as $110\%$. The sensitivity of the magnetometer with NV center ensemble is experimentally improved by a factor of $4$, comparing to dynamical decoupling with normal rectangular pulse. Our work marks an important step towards high trustworthy ultra-sensitive quantum sensing with imperfect quantum control in realistic systems. The used principle however, is universal and not restricted to NV center ensemble magnetometer.}
\end{abstract}
\maketitle
\newcommand{\upcitep}[1]{\textsuperscript{\textsuperscript{\citep{#1}}}}
\citestyle{nature}
\maketitle
\vspace{5mm}
\leftline{\textsf{\textbf{INTRODUCTION}}}\par
\vspace{1mm}
\setlength\parindent{0em}
\textsf{Solid-state electronic spin defects, including nitrogen-vacancy (NV)\cite{awschalom2018quantum}, silicon-vacancy\cite{rose2018observation} and germanium-vacancy\cite{PhysRevLett.118.223603} centers in diamond, have garnered increasing relevance for quantum science and modern physics metrology. Specially, NV centers have been systematically studied and employed in interdisciplinary applications facilitated by long electron spin coherence time\cite{balasubramanian2009ultralong,dong2016reviving} at room-temperature, as well as optically detected magnetic resonance (ODMR)\cite{dong2018non}. Although, single NV center allows for high-spatial-resolution field imaging\cite{chen2015subdiffraction,PhysRevApplied.6.024019}, the acquisition time for magnetic field imaging is too long. Enhancing sensitivity and reducing acquisition time can be achieved by using ensembles of NV centers. The collected photon count is then magnified by the number $N$ of sensing electron spins. Therefore, the shot-noise limited magnetic field sensitivity is improved by a factor\cite{PhysRevX.5.041001,dong2016reviving} $1/\sqrt N$  with same acquisition time. There have been many practical applications utilizing dense NV center ensembles for high-sensitivity magnetic-field sensing\cite{PhysRevX.5.041001} and wide-field magnetic imaging\cite{wojciechowski2018contributed}, such as measurements of single-neuron action potential\cite{barry2016optical}, current flow in graphene\cite{tetienne2017quantum} and chemical shift spectra from small molecules\cite{glenn2018high}.}

\begin{figure*}[tbp]
\centering
\textsf{\includegraphics[width=16cm]{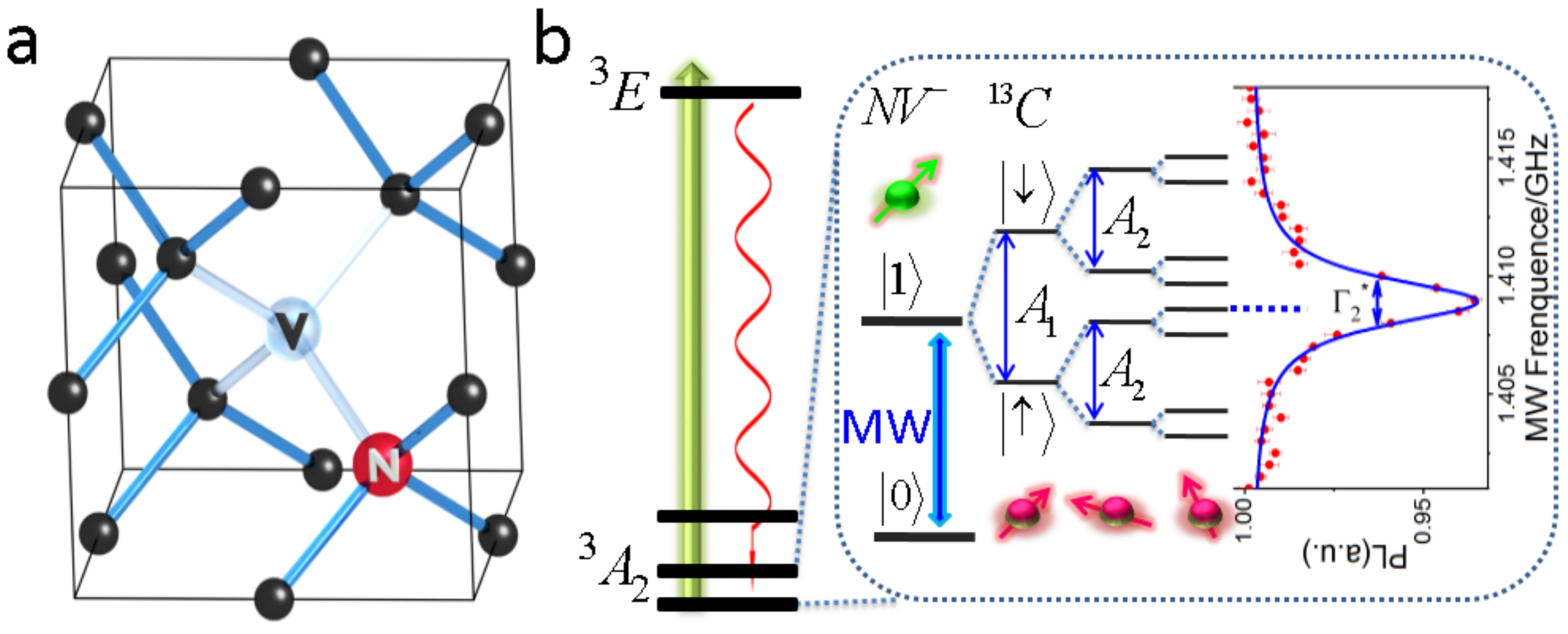}}
\caption{\textsf{(a) Schematic physical structure of the NV center. (b) Energy levels and hyperfine structure of NV center interacting with nearby $^{13}C$ nuclear spins. Laser pulses are used for spin state initialization and readout. MW pulses are used for coherent control of the electron spin. Energy levels ${m_s} = 0$ and ${m_s} = 1$ form sensor qubit in our experiment. Hyperfine sublevels linked to the intrinsic $^{14}N$ nuclear spins of the NV center and P1 centers are not shown. The right panel shows a typical ODMR signal of ${m_s} = 0\left( {\left| 0 \right\rangle } \right)$ and ${m_s} = 1\left( {\left| 1 \right\rangle } \right)$ electron spin sublevels with a $52 mT$ magnetic field along one of four axes of NV center ensemble.}}
\label{fig1}
\end{figure*}

\setlength\parindent{1em}
\textsf{However, solid-state spins always interact with local spin environment, which can induce dephasing and inhomogeneous broadening and inevitably reduce the fidelity of quantum control and sensitivity of quantum sensing. For magnetometry based on NV center ensembles as shown in Fig.\ref{fig1}a, the detection sensitivity is usually limited by quantum dephasing process of electron spin\cite{PhysRevLett.118.093601,PhysRevLett.113.137601,nizovtsev2018non,PhysRevX.8.031025}. The typical electron spin dephasing time is less than $1 \mu s$, which is primarily caused by the detuning ($\delta$) due to inherent or extrinsic inhomogeneous broadening\cite{PhysRevLett.113.137601,nizovtsev2018non,PhysRevX.8.031025,PhysRevB.85.115303,PhysRevB.88.075206,PhysRevLett.115.190801}. There are some inhomogeneous broadening factors contributing to NV centers dephasing process. First, the interaction with randomly localized ${^{13}C}$ nuclei spins with as shown in Fig.\ref{fig1}b, which come from diamond lattice, limits the experimental sensing time to the electron spin dephasing time $T_2^*{\{ ^{13}}C\}  \approx 3\mu s$\cite{dong2018non,PhysRevB.85.115303} with natural isotopic abundance of $1.07\% $ in diamond. Second, the interaction with paramagnetic substitutional nitrogen impurities (P1 centers), which come from nitrogen implantation\cite{dong2018non} or the diamond production process\cite{rondin2014magnetometry} can be a considerable factor of NV center dephasing.
At last, other unknown spins, strain gradients, magnetic-field gradients, and temperature fluctuations contribute to NV center dephasing process.}

\setlength\parindent{1em}
\textsf{Usually, by employing the dynamical decoupling (DD) method or dynamically corrected gates, the spectrum broadening of single NV center can be suppressed to the order in $\delta/\Omega_0 $, where $\Omega_0 $ is the Rabi frequency corresponding to the intensity of control field, in order to achieve fault-tolerant quantum computation\cite{wang2012composite,PhysRevLett.112.050503,PhysRevLett.112.050502,PhysRevLett.115.110502,rong2015experimental}. These methods are valid only if the condition $\delta/\Omega_0\ll1$ is satisfied\cite{wang2012composite,PhysRevLett.112.050503,PhysRevLett.112.050502,PhysRevLett.115.110502}. However, for NV center ensembles with large spectrum broadening, it is a challenge to achieve above condition in practical applications.
The method may make less effective or even create spurious signal\cite{loretz2014single,PhysRevX.5.021009,lang2018non} for large sensing volume\cite{PhysRevLett.120.243604,PhysRevLett.115.110502}. And in fact, the physical criteria for quantum sensing are quite different with quantum computation, which only require\cite{RevModPhys.89.035002}: (1) the quantum system with discrete, resolvable energy levels; (2) quantum state initialization and readout; (3) coherent operation; (4) interaction with a relevant physical field. Hence, on the spirit of modern quantum sensing, we show an effective dynamically composite-pulse operation\cite{wang2012composite,aiello2013composite,PhysRevLett.112.050503} capable of achieving extensive robustness and high sensitive magnetometry based on NV center ensembles in diamond without the condition $\delta/\Omega_0  \ll 1$. By making use of gradient ascent algorithm\cite{qian1999momentum,dolde2014high,rong2015experimental}, we parameterize the phase and duration of control pulse and optimize them with employing experimental testing to meet high fidelity and robustness requirements. By carrying out optimized composite-pulse sequences on NV center ensembles, we can keep almost same as perfect operation for a normalized detuning as large as $110\%$ and improve the sensitivity by a factor of $4$ comparing to dynamical decoupling with normal rectangular pulse. A value of $8$ $nT \cdot H{z^{ - 1/2}} \cdot \mu {m^{3/2}}$ in sensitivity is achieved on NV center ensemble. Hence, we demonstrate a simple and robustness method for wide-field NV centers ensembles quantum magnetometer and outline a path towards effective and trustworthy quantum sensing with quantum optimal control.}

\vspace{3mm}
\leftline{\textbf{\textsf{Sensor spin in diamond and coherent control system}}}\par
\vspace{3mm}
\textsf{The NV center consists of a substitutional nitrogen atom adjacent to a
carbon vacancy in diamond crystal lattice, as shown in Fig.\ref{fig1}a. The ground state of NV center is an electron spin triplet state $^3{A_2}$ and the zero-field splitting between
the ${m_{s}}=0$ and degenerate ${m_{s}}=\pm 1$ sub-levels is $D\approx 2.87$
$GHz$. With secular approximation, the effective
Hamiltonian for NV center of the ground state in a magnetic field $B_z$ along the axis of NV center reads
$H = DS_z^2 - {\gamma _e}{B_z}{S_z}$,
where ${\gamma _{e}}$ is the electron gyromagnetic
ratio, and ${S_{z}}$ is the spin operators of a spin-$1$ system.
The spin-dependent photon
luminescence (PL) enables the implementation of ODMR techniques to detect the spin state. In experiment, we apply a magnetic field at $52 mT$ along the NV symmetry axis to split the  energy levels, and polarize the neighboring nuclear spin to enhance the signal contrast\cite{dong2018non,PhysRevX.5.041001,epstein2005anisotropic}. Under this large magnetic field, the homogeneous broadening effect of strain or electric field can be suppressed\cite{dolde2011electric,PhysRevLett.115.087602,PhysRevX.8.031025}. }

\textsf{When we apply single microwave (MW) to address the transition $\left| {{m_s} = 0} \right\rangle  \leftrightarrow \left| {{m_s} = 1} \right\rangle$, the Hamiltonian of NV center can be expressed as $H = \Omega_0 \vec{n} \cdot \vec{\sigma} + {H_{SB}} + {H_S}$, where $\vec{\sigma}  = ({\sigma _x},{\sigma _y},{\sigma _z})$, $\vec{n} = (\cos\varphi ,\sin\varphi ,0)$ is the effective rotation angle with respect to the x-axis, and $H_{S}$ is the dark spin bath Hamiltonian. Here, due to large energy mismatching, the term of inhomogeneous broadening can be expressed as ${H_{SB}} = {\sigma _z}\sum\nolimits_i {{A_i}\sigma _z^i} $, where $\sigma _z^i$ is the spin operator of the $i$th spin and $A_k$ is the strength of the hyperfine interaction between the sensor qubit and the $i$th spin. The hyperfine interaction with dark spin results in a random local magnetic filed $\delta  = \sum\nolimits_i {{A_i}\sigma _z^i}$ with typical linewidth $\Gamma _2^* \approx  2 MHz$ for our ensemble NV centers as shown in Fig.\ref{fig1}b. Because the typical coherent control is much faster than dynamical fluctuation, we take $\delta$ as a random time independent variable\cite{yang2016quantum,dong2016reviving,wang2012composite,PhysRevLett.112.050503,PhysRevX.8.031025}. Furthermore, we assume that the NV spin ensemble dephasing mechanisms are independent and inhomogeneous broadening effect can be described by a Gaussian linetype ${P(\delta )}$\cite{balasubramanian2009ultralong,rong2015experimental}. }

\begin{figure*}[tbp]
\centering
\textsf{\includegraphics[width=16cm]{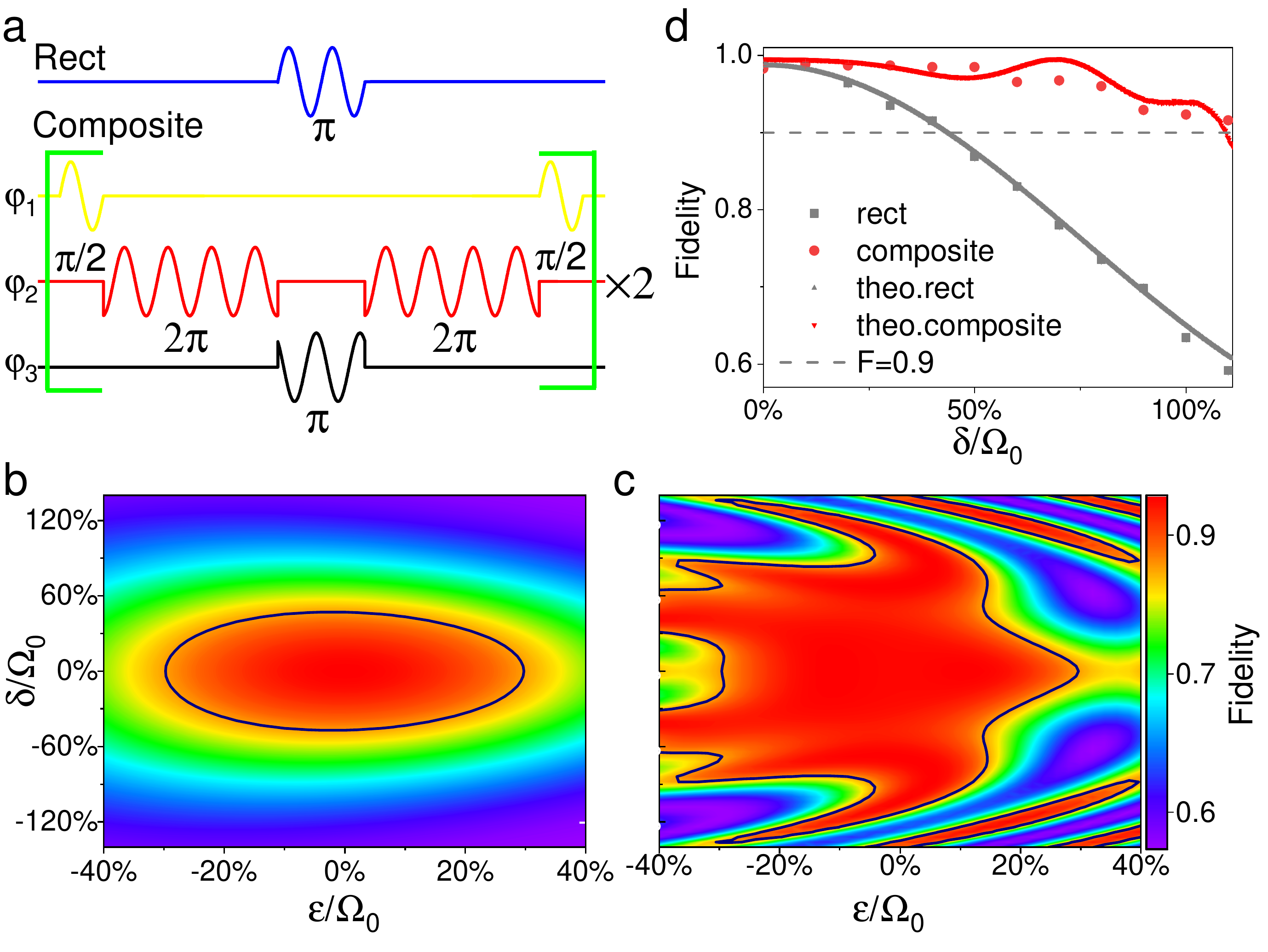}}
\caption{\textsf{(a) General rectangular pulse and composite dynamical sequence for $\pi$ pulse. For the latter, $R(\pi ) = {R^2}\left( {\frac{\pi }{2}} \right) = {\left[ {{R_{{\varphi _1}}}\left( {\frac{\pi }{2}} \right){R_{{\varphi _2}}}\left( {2\pi } \right){R_{{\varphi _3}}}\left( \pi  \right){R_{{\varphi _2}}}\left( {2\pi } \right){R_{{\varphi _1}}}\left( {\frac{\pi }{2}} \right)} \right]^2}$. The relative phases of each MW channel are ${\varphi _2} - {\varphi _1} = {97.08^0},{\varphi _3} - {\varphi _1} =  - {47.88^0}$. (b)-(c) Simulated quantum control fidelity of rectangular (b) and composite dynamical (c) pulses for a range of detuning and control amplitude, scalings with Rabi frequency. The navy lines are contour lines at a fidelity of $0.9$. (d) Measured (dots) and simulated (solid lines) fidelity of $\pi$ pulse. The red (gray) dots correspond to composite (rectangular) pulses.}}
\label{fig2}
\end{figure*}

\vspace{3mm}
\leftline{\textbf{\textsf{Dynamically composite-pulses and robust NV center ensemble
magnetometry}}}
\vspace{3mm}
\textsf{Since the fluctuation of MW's power is always smaller than inhomogeneous broadening in experiment\cite{PhysRevLett.112.050503,rong2015experimental}, we focus on the fidelity optimization of coherent operation with large detuning. More explicitly, we take the target function\cite{PhysRevLett.112.050503,rong2015experimental,nielsen2002simple}
$F(\xi ,U) = \int {d\psi \left\langle \psi  \right|} {U^\dag }\xi \left| \psi  \right\rangle \left\langle \psi  \right|U\left| \psi  \right\rangle, $ where $\xi$ and $U$ are the practical and ideal operations, respectively. The integral is over the uniform measure ${d\psi }$ on state space, normalized by $\int {d\psi }  = 1$. The practical quantum operation can be written as
$\xi \left( {\left| \psi  \right\rangle \left\langle \psi  \right|} \right) = \int {P(\delta )G(\varepsilon ){U_{seq}}(\delta ,\varepsilon )\left| \psi  \right\rangle \left\langle \psi  \right|U_{seq}^\dag d\delta } d\varepsilon, $
where ${G(\varepsilon )}$ is the fluctuation of MW's power and we assume $\varepsilon$ satisfies a Lorentzian distribution\cite{rong2015experimental}. For single qubit operation, the average gate fidelity can be simplified to \cite{bowdrey2002fidelity,rong2015experimental}
\begin{equation}
F(\xi ,U) = \frac{1}{2} + \frac{1}{{12}}\sum\nolimits_{i = x,y,z} {Tr\left( {U{\sigma _i}U\xi \left( {{\sigma _i}} \right)} \right)} \text{.}
\end{equation}
Usually, $\frac{\pi }{2}$ and $\pi$
pulses are employed to form quantum sensing protocol with NV center\cite{RevModPhys.89.035002,RevModPhys.88.041001,dong2016reviving}. Furthermore, the $\pi$ pulse, which acts as the core of DD method, directly affects the final magnetometry sensitivity of NV center\cite{PhysRevX.5.021009,boss2017quantum}.  Based on the dynamical correct core of quantum control\cite{wimperis1994broadband,wang2012composite,rong2015experimental}, we design an effective composite-$\pi$-pulse with two $5$-piece pulses, where relative phases are shown in Fig.\ref{fig2}a. A typical result with such control composite-pulses is shown in Fig.\ref{fig2}c with gradient ascent algorithm. For comparison, we also perform a simulation with a rectangular $\pi$-pulse of the same Rabi frequency (${\Omega _0}$) as shown in Fig.\ref{fig2}b. It is obvious that the demanding robustness requirements for NV center ensemble magnetometry are met. The results greatly enhance the coherent operation fidelity over a large range of detuning and control amplitudes. Then we apply the pulse sequence on a single NV center with a dephasing time of $T_2^{\text{*}} \approx 4$ $\mu s$. As shown in Fig.\ref{fig2}d, we find good agreement with experimental results as theoretical results. The little divergence between theory and experiment is caused by MW's power or relative phase drift. The realistic fidelity can keep $0.9$ with detuning as much as $ \pm {\text{110\% }}$ of the Rabi frequency, which is much superior to rectangular result. Such a control composite-pulse forms the core of our quantum sensing protocol of NV center.}

\begin{figure*}[tbp]
\centering
\textsf{\includegraphics[width=16cm]{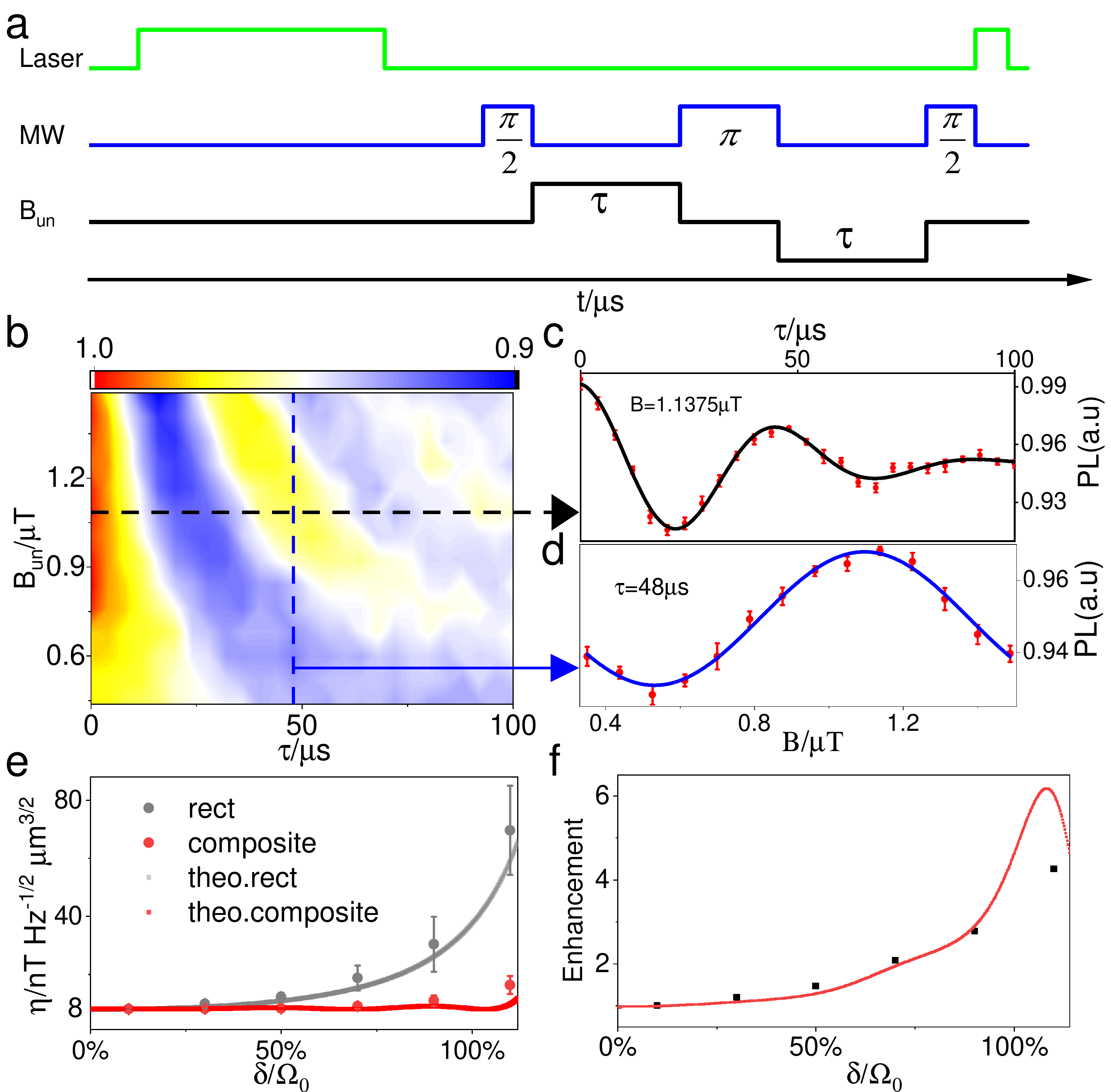}}
\caption{\textsf{(a) AC magnetic-field measurement scheme with pulsed sensing
readout. The NV center ensemble is initialized by $12 mW$ $532nm$ laser pulse with $4$ $\mu {\text{s}}$ duration and then driven by a quantum lock-in amplifier sequence with fixed free precession time $\tau $ to measure an unknown magnetic field.
(b) Results of the spin-echo sensing sequence as
a function of sensing time and the amplitude of AC magnetic field. (c) Spin-echo sequence with fixed amplitude of AC magnetic field. The dephasing time of NV center ensemble is ${T_2} = 104(5)$ $\mu s$ by fitting the signal. (d) Spin-echo sensing sequence as a function of the amplitude of AC magnetic field with fixed sensing time. (e) Measured NV center ensemble-based sensitivity times root sensing volume of a spin-echo magnetometry protocol using rectangular (grey) and composite pulse (red). The grey and green solid lines are theoretical simulation results. (f) The relative sensitivity enhancement of composite-pulse versus rectangular pulse sequences.}}
\label{fig3}
\end{figure*}

\textsf{Based on the high fidelity and robustness of dynamical composite-pulse control on single NV center, we now employ them with an ensemble of inhomogeneously broadened NV center for quantum magnetometry. The NV center ensemble in diamond grown by plasma assisted chemical vapor deposition (CVD) consists of a [N]$ \approx $2 ppm and [NV$^-$]$ \approx 0.0063 ppm$. We perform AC magnetic field sensing protocol by employing DD method\cite{dong2016reviving,PhysRevX.5.041001,RevModPhys.89.035002,RevModPhys.88.041001}. The quantum lock-in sequence is shown in Fig.\ref{fig3}a. To characterize the behavior of such a sensor, we also apply a square wave magnetic field in phase with the spin-echo sensor sequence. The typical spin-echo magnetometry results with rectangular pulse are measured by changing the amplitude of AC magnetic field and sensing time as shown in Fig.\ref{fig3}b without detuning. Under a fixed unknown AC magnetic field, the accumulated relative phase increases linearly with the sensing time and the signal contrast decreases because of inhomogeneous broadening as shown in Fig.\ref{fig3}c. For a fixed sensing time, the accumulated phase has a linear relationship with the amplitude of AC magnetic field as shown in Fig.\ref{fig3}d. In general, the sensitivity\cite{li2018enhancing} of a magnetic field measurement is given by $\eta  = \sigma /(dS/d{B_{un}})$, where the standard deviation of the sensors signal $\sigma $ is compared to the response of the system $dS$ in a changing magnetic field $d{B_{un}}$ and can be calculated by fitting the curve in Fig.\ref{fig3}b or Fig.\ref{fig3}c.}

\textsf{In order to emulate conditions typically for wide-field magnetic field imaging case, we scan the detuning of our NV center ensemble across a large range of values. And for comparison, the final results are shown in Fig.\ref{fig3}e. We can see the sensitivity with composite-pulses keeps almost constant up to $110\%$ detuning and agrees with theoretical prediction. However, for the rectangular pulses, it degenerates quickly when the detuning increases over $50\%$ and the uncertainty of sensitivity is increasing sharply. As detuning further increases for rectangular case, the performance of NV center ensemble is reduced and the fluctuation of sensitivity becomes large. With optimal composite-pulse control, we obtain a stable magnetometric sensitivity times root sensing volume of $8$ $nTH{z^{ - 1/2}}\mu {m^{3/2}}$ for a larger detuning, which is limited by the density and coherence time of NV centers in our diamond sample. It is capable for the investigation of single-neuron action potential\cite{barry2016optical}. When the detuning increases as much as $110\%$ of the resonant Rabi frequency, we experimentally improve the sensitivity by a factor of approximately $4$, comparing to the control with normal rectangular pulse, as shown in Fig.\ref{fig3}f.}

\vspace{3mm}
\leftline{\textbf{\textsf{Discussion}}}\par
\vspace{3mm}

\begin{figure*}[tbp]
\centering
\textsf{\includegraphics[width=12cm]{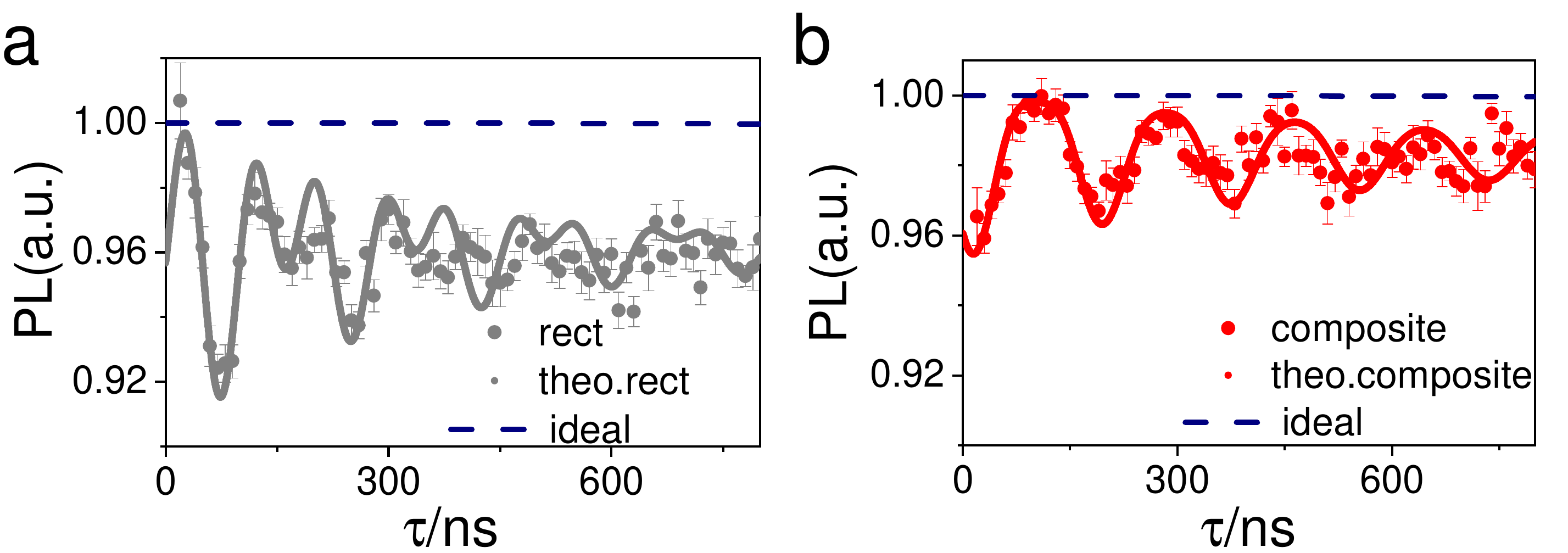}}
\caption{\textsf{(a)-(b) The dynamical process of spin-echo sensing protocol with $110\%$ detuning with rectangular (a) and composite dynamical (b) pulses. The red and grey dots are experimental data. The red and grey curves are theoretical simulation results. The navy dash line is ideal result for spin-echo sensor without detuning and dechoerence.}}
\label{fig4}
\end{figure*}

\textsf{To figure out the physical mechanisms behind of the enhancement for our composite-pulse, we investigate the quantum dynamically  temporal evolution process of our sensing protocol in both theory and experiment as shown in Fig.\ref{fig4}a-b. The sensing signal can be expressed as\cite{dong2016reviving} $s = Tr\left[ {P{e^{ - i{H_2}\tau /\hbar }}R(\pi ){e^{ - i{H_1}\tau /\hbar }}\rho {e^{i{H_1}\tau /\hbar }}R(\pi ) ^\dag {e^{i{H_2}\tau /\hbar }}} \right],$ where the $\rho  = {\rho _{NV}} \otimes {\rho _d}$ is the probe state, ${H_1}$(${H_2}$) is system Hamiltonian under an unknown AC magnetic field, $R(\pi )$ is $\pi$ operation of pulse sequence and $P$ is projection measurement operator. The coherent MW operation $R(\theta )$ can be always expressed by linear combination of the Pauli matrix of spin $1/2$ like $R(\theta ) \triangleq \cos \frac{\theta }{2}I - i\sin\frac{\theta }{2}\vec n \cdot \vec \sigma $ and ${\vec n}$ is the effective rotation axis. So we have
\begin{equation}
s = C + E\operatorname{Re} \left[ {Tr\left( {{U_0}{U_{ 1}}{\rho _d}U_0^\dag U_{ 1}^\dag } \right)} \right]\cos (2\gamma {B_{un}}\tau  + {\phi _1}) + M \text{,}
\end{equation}
where the modulation caused by detuning can be written as ${{M}} = D\operatorname{Re} \left[ {Tr\left( {U_0^2{\rho _d}U_{ 1}^{\dag 2}} \right)} \right]\cos (2\delta \tau  + {\phi _3}) + \operatorname{Re} \left[ {Tr\left( {{U_0}{\rho _d}{U_{ 1}}} \right)} \right]\left[ {A\cos ((\delta  + \gamma {B_{un}})\tau  + {\phi _1}) + B\cos ((\delta  - \gamma {B_{un}})\tau  + {\phi _2})} \right]$ and ${{{U}}_0}$(${{{U}}_{ 1}}$) is the dark spin
propagator conditioned on the NV electron spin state\cite{PhysRevLett.118.150504}. $A$, $B$, $C$, $D$, and $E$ are constant coefficients. Hence, the amplitude of modulation caused by detuning effect decays with the time scale of $T_2^*$. However, the sensing signal decays with the time scale of $T_2$ due to DD method. Due to immune to detuning, the modulation term of composite-pulse sequence is quite small and perfectly canceled out by high fidelity of $\pi$-pulse operation as shown in Fig.\ref{fig4}b. But for the rectangular pulse, the fidelity of $\pi$-pulse operation degenerates seriously with detuning when the $\delta /{\Omega _0} \geqslant 50\% $ as shown in Fig.\ref{fig2}d. The modulation term of detuning effect is significant and decays fast with the time scale of $T_2^*$ as shown in Fig.\ref{fig4}a. So the sensitivity of rectangular pulse sequence degenerates quickly in the signal contrast for detuning effect as shown in Fig.\ref{fig3}e.}

\begin{figure*}[tbp]
\centering
\textsf{\includegraphics[width=12cm]{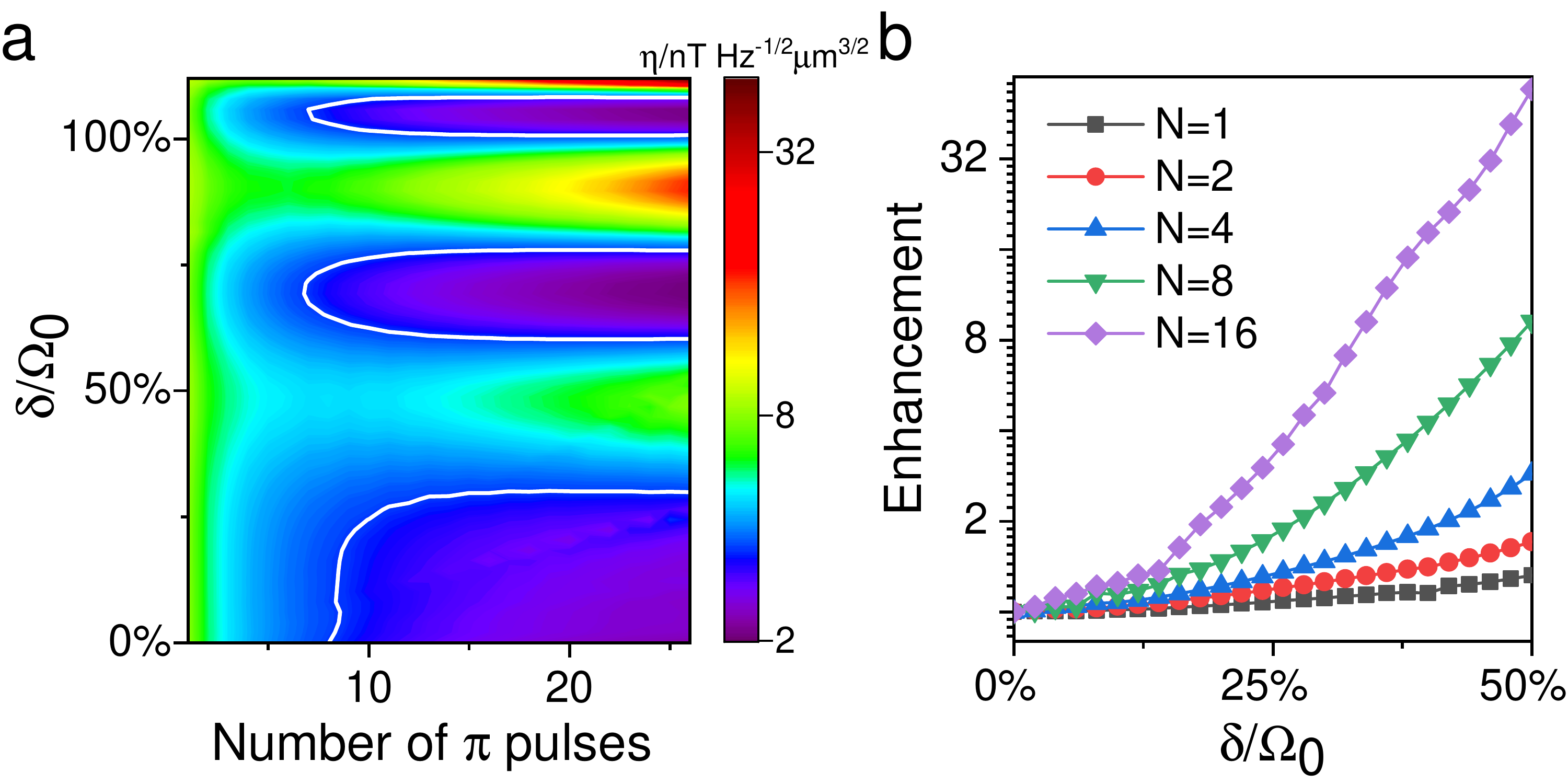}}
\caption{\textsf{(a) Improved magnetometric sensitivity under CPMG sequence for dynamical composite-pulse.
The white lines are
contour lines at sensitivity $4$ $nTH{z^{ - 1/2}}\mu {m^{ 3/2}}$. (b) The relative sensitivity enhancement of composite-pulse versus rectangular pulse sequences for CPMG-N sequences.}}
\label{fig5}
\end{figure*}

\textsf{In this work, we demonstrate how a high magnetic sensitivity in NV center ensemble magnetometry can be maintained over large inhomogeneous broadening with robust quantum optimal control. And our composite pulses sequence, which is simply and effective, enhance the sensitivity up to $4$ comparing with rectangular case under the $110\%$ detuning. In our experiment, we note that the coherence time of NV center ensemble is ${T_2} \approx 100$ $\mu s$, which is mainly limited by P1 center in diamond\cite{PhysRevX.8.031025,PhysRevLett.118.167204}. However, with more advanced quantum control of P1 center\cite{PhysRevX.8.031025}, the coherence time of NV center ensemble can be extended to ${T_2} \approx 1 ms$ for samples with P1 center concentration $ \sim 1 ppm$ with cascaded dynamical decoupling\cite{PhysRevB.86.045214}, which is the limit set by longitudinal spin relaxation of NV centers at room temperature. We also evaluate the Carr-Purcell-Meiboom-Gill (CPMG) sequence for magnetic field sensing as shown in Fig.\ref{fig5}a. The sensitivity can be further improved by a factor of $2$ when the number $\pi$ pulse is greater than $8$. And the relative magnetometry sensitivity enhancement between composite and rectangular sequence increases sharply with the number $\pi$ pulse with the detuning as shown in Fig.\ref{fig5}b. From the theoretical simulation, the performance of the composite-pulse cooperating with CPMG is much better than that of rectangular pulse under large detuning, corresponding to large spectrum broadening in practical applications.}

\textsf{Furthermore, the generating efficiency of NV center ensemble is less than $1\%$ for our diamond sample. And by electron irradiation treatment, the density of NV center can be enhanced more than $30$ times\cite{PhysRevX.8.031025}, which boosts magnetic sensitivity of our NV center ensemble towards sub-$1$ $nTH{z^{ - 1/2}}\mu {m^{3/2}}$. Hence, based on the physical criteria of quantum sensing, we provide a clear path towards effective and reliable quantum sensing with quantum optimal control, which can also be applied to other quantum systems, such as quantum dots, superconducting systems and trapped ions. By combining with large area ring-shaped resonators\cite{PhysRevLett.120.243604}, the magnetometry based on NV center ensemble allows for detecting proton spins in water in a microscopically resolvable volume, bringing new fields of application into reach: for example, noninvasive \emph{in vivo} sensing of biomagnetic fields and microfluidic chemical analysis, as well as cellular and neuronal action imaging\cite{barry2016optical}, where characteristic dimensions are on the micron scale.}


\begin{thebibliography}{10}
\expandafter\ifx\csname url\endcsname\relax
  \def\url#1{\texttt{#1}}\fi
\expandafter\ifx\csname urlprefix\endcsname\relax\def\urlprefix{URL }\fi
\providecommand{\bibinfo}[2]{#2}
\providecommand{\eprint}[2][]{\url{#2}}

\bibitem{awschalom2018quantum}
\bibinfo{author}{Awschalom, D.~D.}, \bibinfo{author}{Hanson, R.},
  \bibinfo{author}{Wrachtrup, J.} \& \bibinfo{author}{Zhou, B.~B.}
\newblock \bibinfo{title}{Quantum technologies with optically interfaced
  solid-state spins}.
\newblock \emph{\bibinfo{journal}{Nat. Photon.}} \textbf{\bibinfo{volume}{12}},
  \bibinfo{pages}{516} (\bibinfo{year}{2018}).

\bibitem{rose2018observation}
\bibinfo{author}{Rose, B.~C.} \emph{et~al.}
\newblock \bibinfo{title}{Observation of an environmentally insensitive
  solid-state spin defect in diamond}.
\newblock \emph{\bibinfo{journal}{Science}} \textbf{\bibinfo{volume}{361}},
  \bibinfo{pages}{60--63} (\bibinfo{year}{2018}).

\bibitem{PhysRevLett.118.223603}
\bibinfo{author}{Bhaskar, M.~K.} \emph{et~al.}
\newblock \bibinfo{title}{Quantum nonlinear optics with a germanium-vacancy
  color center in a nanoscale diamond waveguide}.
\newblock \emph{\bibinfo{journal}{Phys. Rev. Lett.}}
  \textbf{\bibinfo{volume}{118}}, \bibinfo{pages}{223603}
  (\bibinfo{year}{2017}).

\bibitem{balasubramanian2009ultralong}
\bibinfo{author}{Balasubramanian, G.} \emph{et~al.}
\newblock \bibinfo{title}{Ultralong spin coherence time in isotopically
  engineered diamond}.
\newblock \emph{\bibinfo{journal}{Nat. Mater.}} \textbf{\bibinfo{volume}{8}},
  \bibinfo{pages}{383} (\bibinfo{year}{2009}).

\bibitem{dong2016reviving}
\bibinfo{author}{Dong, Y.}, \bibinfo{author}{Chen, X.-D.},
  \bibinfo{author}{Guo, G.-C.} \& \bibinfo{author}{Sun, F.-W.}
\newblock \bibinfo{title}{Reviving the precision of multiple entangled probes
  in an open system by simple $\pi$-pulse sequences}.
\newblock \emph{\bibinfo{journal}{Phys. Rev. A}} \textbf{\bibinfo{volume}{94}},
  \bibinfo{pages}{052322} (\bibinfo{year}{2016}).

\bibitem{dong2018non}
\bibinfo{author}{Dong, Y.} \emph{et~al.}
\newblock \bibinfo{title}{Non-markovianity-assisted high-fidelity
  Deutsch-Jozsa algorithm in diamond}.
\newblock \emph{\bibinfo{journal}{npj Quantum Inf.}}
  \textbf{\bibinfo{volume}{4}}, \bibinfo{pages}{3} (\bibinfo{year}{2018}).

\bibitem{chen2015subdiffraction}
\bibinfo{author}{Chen, X.} \emph{et~al.}
\newblock \bibinfo{title}{Subdiffraction optical manipulation of the charge
  state of nitrogen vacancy center in diamond}.
\newblock \emph{\bibinfo{journal}{Light: Sci. Appl.}}
  \textbf{\bibinfo{volume}{4}}, \bibinfo{pages}{e230} (\bibinfo{year}{2015}).

\bibitem{PhysRevApplied.6.024019}
\bibinfo{author}{Ma, W.-L.} \& \bibinfo{author}{Liu, R.-B.}
\newblock \bibinfo{title}{Angstrom-resolution magnetic resonance imaging of
  single molecules via wave-function fingerprints of nuclear spins}.
\newblock \emph{\bibinfo{journal}{Phys. Rev. Appl.}}
  \textbf{\bibinfo{volume}{6}}, \bibinfo{pages}{024019} (\bibinfo{year}{2016}).

\bibitem{PhysRevX.5.041001}
\bibinfo{author}{Wolf, T.} \emph{et~al.}
\newblock \bibinfo{title}{Subpicotesla diamond magnetometry}.
\newblock \emph{\bibinfo{journal}{Phys. Rev. X}} \textbf{\bibinfo{volume}{5}},
  \bibinfo{pages}{041001} (\bibinfo{year}{2015}).

\bibitem{wojciechowski2018contributed}
\bibinfo{author}{Wojciechowski, A.~M.} \emph{et~al.}
\newblock \bibinfo{title}{Camera-limits for wide-field
  magnetic resonance imaging with a nitrogen-vacancy spin sensor}.
\newblock \emph{\bibinfo{journal}{Rev. Sci. Instrum.}}
  \textbf{\bibinfo{volume}{89}}, \bibinfo{pages}{031501}
  (\bibinfo{year}{2018}).

\bibitem{barry2016optical}
\bibinfo{author}{Barry, J.~F.} \emph{et~al.}
\newblock \bibinfo{title}{Optical magnetic detection of single-neuron action
  potentials using quantum defects in diamond}.
\newblock \emph{\bibinfo{journal}{Proc. Natl. Acad. Sci. USA}}
  \textbf{\bibinfo{volume}{113}}, \bibinfo{pages}{14133--14138}
  (\bibinfo{year}{2016}).

\bibitem{tetienne2017quantum}
\bibinfo{author}{Tetienne, J.-P.} \emph{et~al.}
\newblock \bibinfo{title}{Quantum imaging of current flow in graphene}.
\newblock \emph{\bibinfo{journal}{Sci. Adv.}} \textbf{\bibinfo{volume}{3}},
  \bibinfo{pages}{e1602429} (\bibinfo{year}{2017}).

\bibitem{glenn2018high}
\bibinfo{author}{Glenn, D.~R.} \emph{et~al.}
\newblock \bibinfo{title}{High-resolution magnetic resonance spectroscopy using
  a solid-state spin sensor}.
\newblock \emph{\bibinfo{journal}{Nature}} \textbf{\bibinfo{volume}{555}},
  \bibinfo{pages}{351} (\bibinfo{year}{2018}).

\bibitem{PhysRevLett.118.093601}
\bibinfo{author}{Choi, J.} \emph{et~al.}
\newblock \bibinfo{title}{Depolarization dynamics in a strongly interacting
  solid-state spin ensemble}.
\newblock \emph{\bibinfo{journal}{Phys. Rev. Lett.}}
  \textbf{\bibinfo{volume}{118}}, \bibinfo{pages}{093601}
  (\bibinfo{year}{2017}).

\bibitem{PhysRevLett.113.137601}
\bibinfo{author}{Dr\'eau, A.} \emph{et~al.}
\newblock \bibinfo{title}{Probing the dynamics of a nuclear spin bath in
  diamond through time-resolved central spin magnetometry}.
\newblock \emph{\bibinfo{journal}{Phys. Rev. Lett.}}
  \textbf{\bibinfo{volume}{113}}, \bibinfo{pages}{137601}
  (\bibinfo{year}{2014}).

\bibitem{nizovtsev2018non}
\bibinfo{author}{Nizovtsev, A.} \emph{et~al.}
\newblock \bibinfo{title}{Non-flipping $^{13}$C spins near an NV center in diamond:
  hyperfine and spatial characteristics by density functional theory simulation
  of the C$_{510}$[NV]H$_{252}$ cluster}.
\newblock \emph{\bibinfo{journal}{New J. Phys.}} \textbf{\bibinfo{volume}{20}},
  \bibinfo{pages}{023022} (\bibinfo{year}{2018}).

\bibitem{PhysRevX.8.031025}
\bibinfo{author}{Bauch, E.} \emph{et~al.}
\newblock \bibinfo{title}{Ultralong dephasing times in solid-state spin
  ensembles via quantum control}.
\newblock \emph{\bibinfo{journal}{Phys. Rev. X}} \textbf{\bibinfo{volume}{8}},
  \bibinfo{pages}{031025} (\bibinfo{year}{2018}).

\bibitem{PhysRevB.85.115303}
\bibinfo{author}{Zhao, N.}, \bibinfo{author}{Ho, S.-W.} \&
  \bibinfo{author}{Liu, R.-B.}
\newblock \bibinfo{title}{Decoherence and dynamical decoupling control of
  nitrogen vacancy center electron spins in nuclear spin baths}.
\newblock \emph{\bibinfo{journal}{Phys. Rev. B}} \textbf{\bibinfo{volume}{85}},
  \bibinfo{pages}{115303} (\bibinfo{year}{2012}).

\bibitem{PhysRevB.88.075206}
\bibinfo{author}{Yamamoto, T.} \emph{et~al.}
\newblock \bibinfo{title}{Extending spin coherence times of diamond qubits by
  high-temperature annealing}.
\newblock \emph{\bibinfo{journal}{Phys. Rev. B}} \textbf{\bibinfo{volume}{88}},
  \bibinfo{pages}{075206} (\bibinfo{year}{2013}).

\bibitem{PhysRevLett.115.190801}
\bibinfo{author}{N\"obauer, T.} \emph{et~al.}
\newblock \bibinfo{title}{Smooth optimal quantum control for robust solid-state
  spin magnetometry}.
\newblock \emph{\bibinfo{journal}{Phys. Rev. Lett.}}
  \textbf{\bibinfo{volume}{115}}, \bibinfo{pages}{190801}
  (\bibinfo{year}{2015}).

\bibitem{rondin2014magnetometry}
\bibinfo{author}{Rondin, L.} \emph{et~al.}
\newblock \bibinfo{title}{Magnetometry with nitrogen-vacancy defects in
  diamond}.
\newblock \emph{\bibinfo{journal}{Rep. Prog. Phys.}}
  \textbf{\bibinfo{volume}{77}}, \bibinfo{pages}{056503}
  (\bibinfo{year}{2014}).

\bibitem{wang2012composite}
\bibinfo{author}{Wang, X.} \emph{et~al.}
\newblock \bibinfo{title}{Composite pulses for robust universal control of
  singlet--triplet qubits}.
\newblock \emph{\bibinfo{journal}{Nat. Commun.}} \textbf{\bibinfo{volume}{3}},
  \bibinfo{pages}{997} (\bibinfo{year}{2012}).

\bibitem{PhysRevLett.112.050503}
\bibinfo{author}{Rong, X.} \emph{et~al.}
\newblock \bibinfo{title}{Implementation of dynamically corrected gates on a
  single electron spin in diamond}.
\newblock \emph{\bibinfo{journal}{Phys. Rev. Lett.}}
  \textbf{\bibinfo{volume}{112}}, \bibinfo{pages}{050503}
  (\bibinfo{year}{2014}).

\bibitem{PhysRevLett.112.050502}
\bibinfo{author}{Zhang, J.}, \bibinfo{author}{Souza, A.~M.},
  \bibinfo{author}{Brandao, F.~D.} \& \bibinfo{author}{Suter, D.}
\newblock \bibinfo{title}{Protected quantum computing: Interleaving gate
  operations with dynamical decoupling sequences}.
\newblock \emph{\bibinfo{journal}{Phys. Rev. Lett.}}
  \textbf{\bibinfo{volume}{112}}, \bibinfo{pages}{050502}
  (\bibinfo{year}{2014}).

\bibitem{PhysRevLett.115.110502}
\bibinfo{author}{Zhang, J.} \& \bibinfo{author}{Suter, D.}
\newblock \bibinfo{title}{Experimental protection of two-qubit quantum gates
  against environmental noise by dynamical decoupling}.
\newblock \emph{\bibinfo{journal}{Phys. Rev. Lett.}}
  \textbf{\bibinfo{volume}{115}}, \bibinfo{pages}{110502}
  (\bibinfo{year}{2015}).

\bibitem{rong2015experimental}
\bibinfo{author}{Rong, X.} \emph{et~al.}
\newblock \bibinfo{title}{Experimental fault-tolerant universal quantum gates
  with solid-state spins under ambient conditions}.
\newblock \emph{\bibinfo{journal}{Nat. Commun.}} \textbf{\bibinfo{volume}{6}},
  \bibinfo{pages}{8748} (\bibinfo{year}{2015}).

\bibitem{loretz2014single}
\bibinfo{author}{Loretz, M.} \emph{et~al.}
\newblock \bibinfo{title}{Retraction: Single-proton spin detection by diamond magnetometry}.
\newblock \emph{\bibinfo{journal}{Science}} \bibinfo{pages}{1259464} (\bibinfo{year}{2014}).

\bibitem{PhysRevX.5.021009}
\bibinfo{author}{Loretz, M.} \emph{et~al.}
\newblock \bibinfo{title}{Spurious harmonic response of multipulse quantum
  sensing sequences}.
\newblock \emph{\bibinfo{journal}{Phys. Rev. X}} \textbf{\bibinfo{volume}{5}},
  \bibinfo{pages}{021009} (\bibinfo{year}{2015}).

\bibitem{lang2018non}
\bibinfo{author}{Lang, J.} \emph{et~al.}
\newblock \bibinfo{title}{The non-vanishing effect of detuning errors in
  dynamical decoupling based quantum sensing experiments}.
\newblock \emph{\bibinfo{journal}{arXiv:1809.03234}}  (\bibinfo{year}{2018}).

\bibitem{PhysRevLett.120.243604}
\bibinfo{author}{Rosenfeld, E.~L.}, \bibinfo{author}{Pham, L.~M.},
  \bibinfo{author}{Lukin, M.~D.} \& \bibinfo{author}{Walsworth, R.~L.}
\newblock \bibinfo{title}{Sensing coherent dynamics of electronic spin clusters
  in solids}.
\newblock \emph{\bibinfo{journal}{Phys. Rev. Lett.}}
  \textbf{\bibinfo{volume}{120}}, \bibinfo{pages}{243604}
  (\bibinfo{year}{2018}).

\bibitem{RevModPhys.89.035002}
\bibinfo{author}{Degen, C.~L.}, \bibinfo{author}{Reinhard, F.} \&
  \bibinfo{author}{Cappellaro, P.}
\newblock \bibinfo{title}{Quantum sensing}.
\newblock \emph{\bibinfo{journal}{Rev. Mod. Phys.}}
  \textbf{\bibinfo{volume}{89}}, \bibinfo{pages}{035002}
  (\bibinfo{year}{2017}).

\bibitem{aiello2013composite}
\bibinfo{author}{Aiello, C.~D.}, \bibinfo{author}{Hirose, M.} \&
  \bibinfo{author}{Cappellaro, P.}
\newblock \bibinfo{title}{Composite-pulse magnetometry with a solid-state
  quantum sensor}.
\newblock \emph{\bibinfo{journal}{Nat. Commun.}} \textbf{\bibinfo{volume}{4}},
  \bibinfo{pages}{1419} (\bibinfo{year}{2013}).

\bibitem{qian1999momentum}
\bibinfo{author}{Qian, N.}
\newblock \bibinfo{title}{On the momentum term in gradient descent learning
  algorithms}.
\newblock \emph{\bibinfo{journal}{Neural Networks}} \textbf{\bibinfo{volume}{12}},
  \bibinfo{pages}{145--151} (\bibinfo{year}{1999}).

\bibitem{dolde2014high}
\bibinfo{author}{Dolde, F.} \emph{et~al.}
\newblock \bibinfo{title}{High-fidelity spin entanglement using optimal
  control}.
\newblock \emph{\bibinfo{journal}{Nat. Commun.}} \textbf{\bibinfo{volume}{5}},
  \bibinfo{pages}{3371} (\bibinfo{year}{2014}).

\bibitem{epstein2005anisotropic}
\bibinfo{author}{Epstein, R.}, \bibinfo{author}{Mendoza, F.},
  \bibinfo{author}{Kato, Y.} \& \bibinfo{author}{Awschalom, D.}
\newblock \bibinfo{title}{Anisotropic interactions of a single spin and
  dark-spin spectroscopy in diamond}.
\newblock \emph{\bibinfo{journal}{Nat. Phys.}} \textbf{\bibinfo{volume}{1}},
  \bibinfo{pages}{94} (\bibinfo{year}{2005}).

\bibitem{dolde2011electric}
\bibinfo{author}{Dolde, F.} \emph{et~al.}
\newblock \bibinfo{title}{Electric-field sensing using single diamond spins}.
\newblock \emph{\bibinfo{journal}{Nat. Phys.}} \textbf{\bibinfo{volume}{7}},
  \bibinfo{pages}{459} (\bibinfo{year}{2011}).

\bibitem{PhysRevLett.115.087602}
\bibinfo{author}{Kim, M.} \emph{et~al.}
\newblock \bibinfo{title}{Decoherence of near-surface nitrogen-vacancy centers
  due to electric field noise}.
\newblock \emph{\bibinfo{journal}{Phys. Rev. Lett.}}
  \textbf{\bibinfo{volume}{115}}, \bibinfo{pages}{087602}
  (\bibinfo{year}{2015}).

\bibitem{yang2016quantum}
\bibinfo{author}{Yang, W.}, \bibinfo{author}{Ma, W.-L.} \&
  \bibinfo{author}{Liu, R.-B.}
\newblock \bibinfo{title}{Quantum many-body theory for electron spin
  decoherence in nanoscale nuclear spin baths}.
\newblock \emph{\bibinfo{journal}{Rep. Prog. Phys.}}
  \textbf{\bibinfo{volume}{80}}, \bibinfo{pages}{016001}
  (\bibinfo{year}{2016}).

\bibitem{nielsen2002simple}
\bibinfo{author}{Nielsen, M.~A.}
\newblock \bibinfo{title}{A simple formula for the average gate fidelity of a
  quantum dynamical operation}.
\newblock \emph{\bibinfo{journal}{Phys. Lett. A}}
  \textbf{\bibinfo{volume}{303}}, \bibinfo{pages}{249--252}
  (\bibinfo{year}{2002}).

\bibitem{bowdrey2002fidelity}
\bibinfo{author}{Bowdrey, M.~D.}, \bibinfo{author}{Oi, D.~K.},
  \bibinfo{author}{Short, A.~J.}, \bibinfo{author}{Banaszek, K.} \&
  \bibinfo{author}{Jones, J.~A.}
\newblock \bibinfo{title}{Fidelity of single qubit maps}.
\newblock \emph{\bibinfo{journal}{Phys. Lett. A}}
  \textbf{\bibinfo{volume}{294}}, \bibinfo{pages}{258--260}
  (\bibinfo{year}{2002}).

\bibitem{RevModPhys.88.041001}
\bibinfo{author}{Suter, D.} \& \bibinfo{author}{\'Alvarez, G.~A.}
\newblock \bibinfo{title}{Colloquium: Protecting quantum information against
  environmental noise}.
\newblock \emph{\bibinfo{journal}{Rev. Mod. Phys.}}
  \textbf{\bibinfo{volume}{88}}, \bibinfo{pages}{041001}
  (\bibinfo{year}{2016}).

\bibitem{boss2017quantum}
\bibinfo{author}{Boss, J.~M.}, \bibinfo{author}{Cujia, K.},
  \bibinfo{author}{Zopes, J.} \& \bibinfo{author}{Degen, C.~L.}
\newblock \bibinfo{title}{Quantum sensing with arbitrary frequency resolution}.
\newblock \emph{\bibinfo{journal}{Science}} \textbf{\bibinfo{volume}{356}},
  \bibinfo{pages}{837--840} (\bibinfo{year}{2017}).

\bibitem{wimperis1994broadband}
\bibinfo{author}{Wimperis, S.}
\newblock \bibinfo{title}{Broadband, narrowband, and passband composite pulses
  for use in advanced nmr experiments}.
\newblock \emph{\bibinfo{journal}{J. Magn. Reson. Ser. A}}
  \textbf{\bibinfo{volume}{109}}, \bibinfo{pages}{221--231}
  (\bibinfo{year}{1994}).

\bibitem{li2018enhancing}
\bibinfo{author}{Li, C.-H.} \emph{et~al.}
\newblock \bibinfo{title}{Enhancing the sensitivity of a single electron spin
  sensor by multi-frequency control}.
\newblock \emph{\bibinfo{journal}{Appl. Phys. Lett.}}
  \textbf{\bibinfo{volume}{113}}, \bibinfo{pages}{072401}
  (\bibinfo{year}{2018}).

\bibitem{PhysRevLett.118.150504}
\bibinfo{author}{Liu, G.-Q.} \emph{et~al.}
\newblock \bibinfo{title}{Single-shot readout of a nuclear spin weakly coupled
  to a nitrogen-vacancy center at room temperature}.
\newblock \emph{\bibinfo{journal}{Phys. Rev. Lett.}}
  \textbf{\bibinfo{volume}{118}}, \bibinfo{pages}{150504}
  (\bibinfo{year}{2017}).

\bibitem{PhysRevLett.118.167204}
\bibinfo{author}{Lillie, S.~E.} \emph{et~al.}
\newblock \bibinfo{title}{Environmentally mediated coherent control of a spin
  qubit in diamond}.
\newblock \emph{\bibinfo{journal}{Phys. Rev. Lett.}}
  \textbf{\bibinfo{volume}{118}}, \bibinfo{pages}{167204}
  (\bibinfo{year}{2017}).

\bibitem{PhysRevB.86.045214}
\bibinfo{author}{Pham, L.~M.} \emph{et~al.}
\newblock \bibinfo{title}{Enhanced solid-state multispin metrology using
  dynamical decoupling}.
\newblock \emph{\bibinfo{journal}{Phys. Rev. B}} \textbf{\bibinfo{volume}{86}},
  \bibinfo{pages}{045214} (\bibinfo{year}{2012}).

\bibitem{ziem2018quantitative}
\bibinfo{author}{Ziem, F.}, \bibinfo{author}{Garsi, M.},
  \bibinfo{author}{Fedder, H.} \& \bibinfo{author}{Wrachtrup, J.}
\newblock \bibinfo{title}{Quantitative nanoscale mri with a wide field of
  view}.
\newblock \emph{\bibinfo{journal}{arXiv:1807.08343}}  (\bibinfo{year}{2018}).

\bibitem{Philipp2018Robust}
\bibinfo{author}{Konzelmann, P.} \emph{et~al.}
\newblock \bibinfo{title}{Robust and efficient quantum optimal control of spin
  probes in a complex (biological) environment}.
\newblock \emph{\bibinfo{journal}{arXiv:1808.0408}}  (\bibinfo{year}{2018}).

\end{thebibliography}

\begin{figure*}[tbp]
\centering
\textsf{\includegraphics[width=16cm]{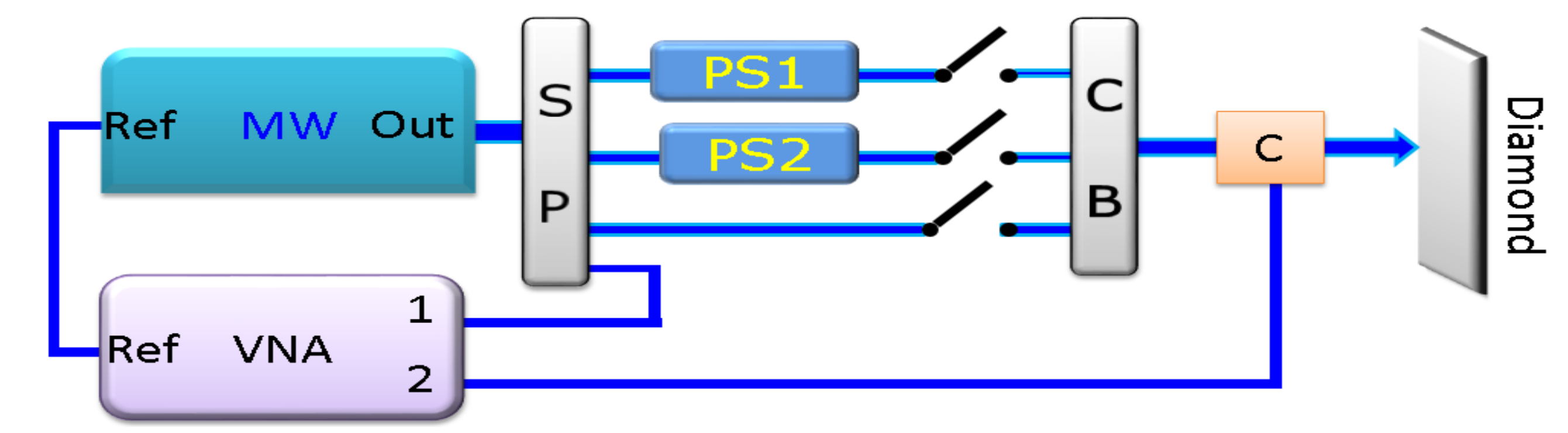}}
\caption{\textsf{MW circuit for dynamical composite-pulse operation experiment. The MW signals are split to four channels and sent to high-frequency switches. Two variable phase shifters (PS, ARRA::N4428D) connected to two of the channels are used to adjust the relative phases. The three channels are combined and sent through circulators (C) to the diamond sample with a copper wire. The phase is measured by beating each of the channels with a proper attenuation on a VNA (ROHDE\&SCHWARZ::ZNB8). The relative phase is calibrated up to ${0.01^0}$.}}
\label{fig6}
\end{figure*}

\vspace{3mm}
\leftline{\textbf{\textsf{Methods}}}
\vspace{3mm}
\textsf{The measurement setup is largely the same as described in the work of Dong,Y et al.\cite{dong2018non}. Here we repeat some parts of that description and explain important differences. In the experiment, we make use of power splitters and phase shifters to control the relative phases between different MW channels. And we calibrate the phase shift with vector network analyze (VNA) and operate the fields using the circuit presented in Fig.\ref{fig6}. For the single NV center experiment, naturally occurring defects in pure CVD diamonds (Element-6) is used \cite{dong2018non}.}

\textsf{Our NV ensemble magnetometry experiments are performed on a CVD grown diamond. The sample is grown on type Ib commercial high pressure high temperature (HPHT) (100)-oriented single crystal diamonds of approximate dimensions ${\text{3}} \times 3 \times 0.5$ ${\text{m}}{{\text{m}}^3}$ purchased from Element-6. Before we use the diamond substrates, they are cleaned in a mixture of sulfuric and nitric acid ($1:2$) for $1 h$ at $200$ $^0{\text{C}}$ and rinsed with de-ionized water, acetone, and isopropanol. Prior to growth, the diamonds substrates are pre-treatment using a ${{\text{H}}_2}$/${{\text{O}}_2}$ plasma in order to prepare their surfaces for single crystal diamond growth. The microwave plasma CVD system (MPCVD, Seki Technotron Corp. AX-5250S) is used to growth the samples. The typical growth parameters are 4.5$\sim$5.0 kW microwave power at a pressure of 140$\sim$160 Torr. The growth temperature is about 850-1000 $^0{\text{C}}$. Hydrogen (5N), $6\% $ ${\text{C}}{{\text{H}}_4}$ (5N) and no intentional addition of nitrogen (${N_2}$) are used to growth. The growth time for the sample is $72$ $h$, and the growth rate is about $20$$\mu m/h$. After growth, the sample is separated from the HPHT diamond substrate by laser cutting, and both sides of the growth plates are polished by mechanical polisher. We study naturally occurring NV centers that are located $5\mu m$ below the surface.}

\vspace{3mm}
\leftline{\textbf{\textsf{Acknowledgements}}}\par

\textsf{This work is supported by The National Key Research and Development Program of China (No. 2017YFA0304504), the National Natural Science
Foundation of China (Nos. 11374290, 61522508, 91536219, and 11504363).}

\textsf{\emph{Note added}.---While finalizing this manuscript we became
aware of two complementary studies that considered
the sensing under large inhomogeneous control fields\cite{ziem2018quantitative} and its application in temperature sensing\cite{Philipp2018Robust} with diamond.}


\clearpage

\end{document}